\documentclass[12pt,preprint,iop,apj]{emulateapj}  
\usepackage{graphicx}
\usepackage{txfonts}
\newcommand{\va}{v_{\mathrm{A}}}

\newcommand{\der}{{\rm d}}

\newcommand{\rhoi}{\rho_{\rm i}}
\newcommand{\rhoe}{\rho_{\rm e}}

\usepackage{natbib}

\begin{document}

\title{Analytic approximate seismology of propagating MHD waves in the solar corona}

	\shorttitle{Seismology of propagating waves}

   \author{M. Goossens$^1$, R. Soler$^{1,2}$, I. Arregui$^{3,4}$, \& J. Terradas$^2$}

   \affil{$^1$Centre for Mathematical Plasma Astrophysics, Department of Mathematics, KU Leuven,
               Celestijnenlaan 200B, 3001 Leuven, Belgium}

 \affil{$^2$Solar Physics Group, Departament de F\'isica, Universitat de les Illes Balears,
               E-07122 Palma de Mallorca, Spain}

\affil{$^3$Instituto de Astrof\'isica de Canarias, V\'ia L\'actea s/n, E-38205 La Laguna, Tenerife, Spain}

\affil{$^4$Departamento de Astrof\'isica, Universidad de La Laguna, E-38205 La Laguna, Tenerife, Spain}

\email{marcel.goossens@wis.kuleuven.be}

  \begin{abstract}

Observations show that propagating magnetohydrodynamic (MHD) waves are ubiquitous in the solar atmosphere. The technique of MHD seismology uses the wave observations combined with MHD wave theory to indirectly infer  physical parameters of the solar atmospheric plasma and magnetic field. Here we present an analytical seismological inversion scheme for propagating MHD waves. This scheme uses in a consistent manner  the observational information on wavelengths  and  damping lengths, along with observed values of periods or phase velocities,   and  is based on  approximate asymptotic expressions for the theoretical values of wavelengths  and  damping lengths. The applicability of the inversion scheme is discussed and an example is given.

  \end{abstract}



\section{Introduction}

  Standing transverse magnetohydrodynamic (MHD) waves in coronal loops were first detected in 1999 by \citet{aschwanden1999} and \citet{nakariakov1999} in observations made with the EUV telescope on board of the Transition Region and Coronal Explorer (TRACE).  Since then the detection of these standing MHD waves  has been confirmed and in addition damped standing MHD waves  have been observed in hot coronal loops by the SUMER instrument on board SOHO. More recent observations of transverse loop oscillations with high spatial and temporal resolutions have been made with instruments on board of STEREO and SDO \citep[see, e.g.,][]{verwichte2009,verwichte2010,aschwanden2011,white2012,wang2012}. The transverse loop oscillations have periods, $T$, of the order of $\simeq 2 -10$ minutes and comparatively short damping times, $\tau_{\rm D}$, of the order of $\simeq 3 -20$ minutes. There is general consensus that the transverse loop oscillations are standing kink MHD waves. Kink refers to the azimuthal wavenumber $m=1$ in a system of cylindrical coordinates with its $z$-axis along the axis of the loop. The MHD waves have to be kink because only for this value of the azimuthal wavenumber  the axis of the loop and the loop as a whole  are displaced.

A possible explanation of the observed rapid damping is resonant absorption. \citet{ruderman2002}  were the first to suggest that the observed rapid damping of the transverse oscillations of coronal loops could be explained by resonant absorption. In the context of the heating of solar plasmas  \citet{hollweg1988} have predicted that oscillations  in coronal loops are to undergo rapid damping. In the same context \citet{goossens1992} derived analytical expressions for the frequency and the damping rate of quasi-modes in static and stationary equilibrium models.     \citet{ruderman2002}  focused on proving the principle of resonant absorption as damping mechanism for the transverse standing MHD waves  in coronal loops and considered one specific numerical example. \citet{goossens2002}  looked at the damping times of 11 loop oscillation events and basically confirmed that resonant absorption can explain the observed damping as suggested by    \citet{ruderman2002}.  \citet{arregui2007} and \citet{goossens2008} showed that resonant absorption of kink MHD waves can explain the observed periods and damping times. 

 \citet{goossens2009}  showed that long-wavelength kink MHD waves are primarily driven by magnetic tension force and hence their behavior is more Alfv\'enic than fast magnetosonic.  The condition of long wavelegth is that the wavelength is much longer than the radius of the tube. \citet{goossens2012} showed that kink MHD waves propagate vorticity and that the fundamental radial modes of kink waves are surface Alfv\'{e}n waves. Note that the term Alfv\'enic was already introduced in the 1970s by \citet{ionson1978} and by \citet{wentzel1979}. The surface Alfv\'{e}n wave character of kink MHD waves was already pointed out by \citet{wentzel1979}. An important property of surface Alfv\'{e}n waves is that in a non-uniform plasma their frequency is in the Alfv\'{e}n continuum so that they undergo continuum damping or resonant absorption \citep[see, e.g.,][]{hasegawa1982,goedbloed1983}.

 Standing MHD waves are relatively rare phenomena  since they need an energetic event like a solar flare or strong vortex shedding in order to get excited. A more recent development over the last five years is that observations  show that  small amplitude propagating transverse MHD waves are almost everywhere in the solar atmosphere. They were first observed with the Coronal Multi-channel Polarimeter (CoMP) instrument by \citet{tomczyk2007} and  subsequently by \citet{tomczyk2009} and others. The physical mechanism of resonant absorption does not make a distinction between standing MHD waves and propagating MHD waves. The physical mechanism is the same and the mathematical analysis is largely the same. For a standing MHD wave the axial wavenumber $k_z$ is specified and the frequency $\omega$  is determined. For a propagating MHD wave the frequency $\omega$ is specified and the wavenumber $k_z$ is computed. The damping time $\tau_{\rm D}$ for standing waves is replaced with the damping length $L_{\rm D}$. In what follows we shall present an inversion scheme for propagating MHD waves that uses observed values of the wavelength and damping length in combination with asymptotic analytic expressions to compute theoretical values of the wavelength and the damping length. The theory of resonant damping for propagating MHD waves has been developed by \citet{terradas2010} and used to explain CoMP observations by \citet{verth2010}. Subsequent extensions to the theory were made by \citet{soler2011a,soler2011b,soler2011c}.

The present investigation is inspired by earlier work on seismic inversion for standing MHD waves in coronal loops. \citet{arregui2007} were the first to carry out a seismological investigation of standing transverse MHD waves of coronal loops that used the observed values of both the periods and the damping times in a consistent way. Their seismic inversion was fully numerical and the theoretical values of the periods and damping times were obtained by numerical  eigenvalue computations for one dimensional cylindrical equilibrium models.  As a result the inversion scheme could turn out to be rather involved.  In a subsequent investigation \citet{goossens2008} used  approximate asymptotic expressions for the period and damping rate to work out an analytical  seismological inversion scheme. This analytical inversion scheme has the big advantage that it is amazingly simple to use when compared to the numerical scheme of \citet{arregui2007}. The downside of this scheme is  that its simplicity might invite us to use it outside the domain of validity of the approximate expressions for the period and damping rate. Analytical schemes are also very useful for parameter inference in the Bayesian framework. The main potential of Bayesian inference is the consistent solution to the inverse problem using the forward model and the available observational information. The unknown parameters are constrained by data and uncertainties are correctly propagated from observed data to inferred parameters. Analytical forward and inverse problems greatly simplify this task, a shown by \citet{arregui2011} using the analytical scheme of \citet{goossens2008} to constrain coronal loop properties. In addition, the inversion scheme has been used beyond the context of coronal loop oscillations. For example, it has been used to perform seismology of thin threads of solar prominences, since the transverse oscillations of both coronal loops and prominence threads can be studied using the same physical model \citep[see details in, e.g.,][]{soler2010,arreguiballester2011,arregui2012}.  \citet{soler2011b} suggested that the inversion scheme of \citet{goossens2008} could be adapted to the case of propagating waves. Here we fully explore the seismic inversion for propagating MHD waves.

\section{Asymptotic analytic expressions for the wavelength.}

In this section we derive an expression for the wavelength of a propagating wave with a given frequency in the thin tube approximation. In the following section we shall derive an expression for the damping length  in the thin tube approximation and thin boundary approximation.  Part of the information in this section is  presented in \citet{terradas2010}. \citet{terradas2010} investigated the damping by resonant absorption of propagating MHD waves. They obtained simple expressions for the wavelength and the damping length in the thin tube approximation and the thin boundary approximation.  In addition they went beyond the thin tube and thin boundary approximation by using numerical resistive calculations. 

The analytical expression that we shall use for the wavelength  is obtained by (i) adopting the thin tube (TT)  approximation for MHD waves and by (ii) modeling the wave guide as  a uniform cylinder  with a straight magnetic field along the  $z$-axis.  For a standing MHD wave the axial wavelength is specified by the dimension of the flux tube and the corresponding frequency is determined by the dispersion relation. In that case  the thin tube approximation means that axial wavelength is much longer than the radius $R$ of the tube so that $k_z R \ll 1$. In case of propagating MHD waves the frequency is specified and the wavelength is determined by the dispersion relation. The thin tube approximation now means that during one period specified by the frequency $\omega$ a signal traveling at the Alfv\'{e}n speed $\va$ can cross the wave guide in the radial direction many times or
\begin{equation}
 \frac{\omega}{\va/R} \ll 1.
\end{equation}
 The TT approximation means the wavelength  is independent of the radius and that effects due to non-zero radius are absent as far as the wavelength  is concerned. The choice of a uniform equilibrium model means that effects due to stratification are absent. The wave guide is modeled as a cylindrical plasma with constant density $\rhoi$ embedded in an external plasma with constant density $\rhoe$. The wave guide is basically a density enhancement with $\rhoe < \rhoi$. The magnetic field is constant and has the same strength both inside and outside the tube.

Our starting point is the well-known thin tube approximation of the dispersion relation for non-axisymmetric  MHD waves on a  uniform cylinder with a straight magnetic field along the $z$-axis, namely
\begin{equation}
\rhoi (\omega^2 - \omega_{\rm A,i}^2) + \rhoe (\omega^2 - \omega_{\rm A,e}^2) = 0.
 \label{TrueDisc1}
\end{equation}
The subscripts `i' and `e' refer to quantities respectively in the
tube and in the external plasma surrounding the tube. $\rho$
is the density. $\omega_{\rm A}$ is the local Alfv\'{e}n frequency and
$\va$ is the local Alfv\'{e}n velocity. They are defined as
\begin{equation}
\omega_{\rm A} = k_z \va, \qquad \va = \frac{B}{\sqrt{\mu_0 \rho}}, \label{AlfvenFr}
\end{equation}
where $k_z$ is the longitudinal wavenumber, $B$ is the magnetic field strength and {\bf $\mu_0$ is the magnetic permeability  of free space}. Note that the dispersion relation  (Equation~(\ref{TrueDisc1})) is independent of the azimuthal wavenumber $m$ so that  Equation~(\ref{TrueDisc1})  applies to all non-axisymmetric MHD waves. Our interest is focused on kink waves which have $m=1$.

For standing waves $k_z$ is specified and the dispersion relation Equation~(\ref{TrueDisc1}) is solved for the frequency $\omega$. For propagating waves we consider waves that are generated at a given position with a real frequency $\omega_{\star}$ and we have to solve Equation~(\ref{TrueDisc1}) for $k_z$. The result is 
\begin{equation}
k_z  =  \omega\sqrt{\frac{\rhoi + \rhoe}{2B^2 / \mu}}  \equiv k_{\star}.
\label{TrueDisc2}
\end{equation}
Let us convert frequencies $\omega$  to periods $T=2\pi/\omega$  and wavenumbers $k_z$ to wavelengths $\lambda=2\pi/k_z$   and rewrite  Equation~(\ref{TrueDisc2}) as
\begin{equation}
\lambda   =  \frac{\sqrt{2} v_{\rm A,i} T}{ A(\zeta)}.
 \label{lambda}
\end{equation}
In Equation~(\ref{lambda}) $\zeta$ is the density contrast, namely
\begin{equation}
\zeta = \frac{\rhoi}{\rhoe} > 1.
\label{zeta}
\end{equation}
The function $A(\zeta)$ is defined in Equation~(4) of \citet{goossens2008} as
\begin{equation}
A(\zeta) = \left( \frac{\zeta +1}{\zeta} \right)^{1/2}. \label{A}
\end{equation}
Equation~(\ref{lambda}) is our first key equation. It expresses a relation of the wavelength  $\lambda $, and the period, $T$, which are two observable quantities, in terms of the internal Alfv\'{e}n velocity, $v_{\rm A,i}$ and the density contrast, $\zeta$, which are two quantities that we aim to determine by seismic inversion. Let us recall that Equation~(\ref{lambda}) has been obtained by use of the TT approximation for a uniform cylindrical tube. Effects from non-zero radius and stratification are absent from Equation~(\ref{lambda}).

Let us now look at Equation~(\ref{lambda}) from the seismic point of view. If we have observed values  of the wavelength,  $\lambda$, and the period, $T$, and we convince ourselves that Equation~(\ref{lambda}) is a good first analytical approximation of the wavelength  then we can invert Equation~(\ref{lambda}) for either $v_{\rm A,i}$ or $\zeta$. Actually, we shall do both. Let us first solve Equation~(\ref{lambda})
for $v_{\rm A,i}$. We prefer to use dimensionless quantities and so we introduce $y_{\star}$ as
\begin{equation}
y_{\star} = \frac{v_{\rm A,i} T}{\lambda}. \label{y}
\end{equation}
From Equation~(\ref{lambda}) we obtain
\begin{equation}
y_{\star}  =  \frac{1}{\sqrt{2}}  A(\zeta). \label{y1}
\end{equation}
Since we have not any information on $\zeta$, it might appear that Equation~(\ref{y1}) is not very helpful. However, closer inspection reveals that it contains  valuable information. First note that for given observed wavelength  $\lambda$, Equation~(\ref{y1}) is a parametric representation of $y_{\star}$ (or equivalently of $v_{\rm A,i}$) in terms of $\zeta$. In order to stress this point we define the function $F_{1, \star}$ by use of the right hand member of Equation~(\ref{y1}) as
\begin{equation}
F_{1, \star} \, : \, [1 , \infty[ \, \rightarrow  \mathbb{R},\;\; \zeta \;
\Rightarrow \; F_{1, \star}(\zeta) =  \frac{1}{\sqrt{2}}A(\zeta). \label{F1}
\end{equation}
Second, note that the function $F_{1, \star}$ is strictly decreasing and in addition note that
\begin{equation}
\mbox{max} (F_{1,\star}) = F_{1,\star}(1) = 1,  \;\; \lim_{\zeta \rightarrow \infty} F_{1, \star} (\zeta) = \frac{1}{\sqrt{2}}.
\end{equation}
This means that
\begin{eqnarray}
\frac{ 1}{\sqrt{ 2} } & \leq &  \;y_{\star}  \leq \;1, \label{y21} \\
\frac{ 1}{ \sqrt{ 2} } \frac{ \lambda}{ T} & \leq & v_{\rm A,i} \leq  \; \frac{ \lambda}{T}.
\label{y2}
\end{eqnarray}
Inequality~(\ref{y2})  tells us that the Alfv\'{e}n velocity is  constrained to a narrow range.  Whatever the density contrast is, the Alfv\'{e}n velocity is in between $\lambda/\sqrt{2}T $ and  $\;  \lambda/T $.  

Let us stress again that the seismic variable $y_{\star}$ contains information of two observables, namely the wavelength, $\lambda$, and the period, $T$. However, we note that the ratio $\lambda/T$ can be expressed as
\begin{equation}
 \frac{\lambda}{T} = v_{\rm ph}, \label{eq:periodvph}
\end{equation}
 where $v_{\rm ph}$ is the phase velocity of the propagating wave. Using Equation~(\ref{eq:periodvph}) the definition of $y_{\star}$ becomes
\begin{equation}
y_{\star} = \frac{v_{\rm A,i}}{v_{\rm ph}}. \label{yvph}
\end{equation}
Therefore, for practical purposes we only need information about one observed quantity, i.e., the phase velocity, $v_{\rm ph}$. Hence Inequality~(\ref{y2}) can be rewritten as
\begin{equation}
 \frac{ 1}{ \sqrt{ 2} } v_{\rm ph}  \leq  v_{\rm A,i} \leq  \; v_{\rm ph}.
\end{equation}

With the help of Inequality~(\ref{y21}) we can refine the definition of $F_{1,\star}$ and replace Equation~(\ref{F1}) with
\begin{equation}
F_{1,\star} \;:\;[1, \; \infty[ \;\rightarrow  ]\frac{1}{\sqrt{2}},\;\;1 ],\;\; \zeta \; \Rightarrow \; F_{1,\star}(\zeta) =\frac{1}{\sqrt{2}} \left(\frac{\zeta + 1}{\zeta} \right)^{1/2}.
\label{F1a}
\end{equation}
Let us now solve Equation~(\ref{y1}) for $\zeta$ and find
\begin{equation}
\zeta = \frac{1}{2 y_{\star}^2 - 1}. \label{zeta1}
\end{equation}
Equation~(\ref{zeta1}) is the twin of Equation~(\ref{y1}). For given observed wavelength,  $\lambda$, and period, $T$, or alternatively for a given observed phase velocity, $v_{\rm ph}$, Equation~(\ref{zeta1}) is a parametric representation of $\zeta$ in terms of $y_{\star}$ (or equivalently in terms
of $v_{\rm A,i}$). In order to stress this point we define the function $G_{1,\star}$ by use of the right hand member of Equation~(\ref{zeta1}) as
\begin{equation}
G_{1,\star} \;:\;]\frac{1}{\sqrt{2}}, \;\; 1]  \;\;\rightarrow \mathbb{R},\;\; y_{\star} \; \Rightarrow \; G_{1,\star}(y_{\star}) = \frac{1}{2 y_{\star}^2 - 1}. \label{G1}
\end{equation}
The function $G_{1,\star}$ is strictly decreasing and
\begin{equation}
G_{1, \star}(1) = 1, \;\;\lim_{y \rightarrow 1/\sqrt{2}} G_{1,\star}(y)= \infty.
\end{equation}
With this information on the function $G_{1,\star}$ we can refine its definition (Equation~(\ref{G1})). Combined with the definition of
$F_{1,\star}$ (Equation~(\ref{F1a})) we obtain the following prescriptions for the functions $F_{1,\star}$ and $G_{1,\star}$, namely
\begin{eqnarray}
F_{1,\star} \;&:&\;[1, \; \infty[ \;\rightarrow  [\frac{1}{\sqrt{2}},\;\;1 ],\;\; \zeta \; \Rightarrow \; F_{1,\star}(\zeta) = \frac{1}{\sqrt{2}} \left(\frac{\zeta + 1}{\zeta} \right)^{1/2}. \nonumber \\\label{F1aG1a1} \\
G_{1,\star} \;&:&\;]\frac{1}{\sqrt{2}}, \;\; 1] \;\;\rightarrow \mathbb{R},\;\; y_{\star} \; \Rightarrow \; G_{1,\star}(y_{\star}) = \frac{1}{2 y_{\star}^2 - 1}.
\label{F1aG1a}
\end{eqnarray}
Of course, $G_{1,\star}$ is the inverse function of $F_{1,\star}$, i.e., $G_{1,\star} = F_{1,\star}^{-1}$ and conversely $G_{1,\star}^{-1} = F_{1,\star}$.

Let us recapitulate what seismic information we have deduced from the observations. The quantities $\zeta$ and $y_{\star}$ are in the following intervals
\begin{equation}
\zeta \in I_{\zeta} = [1,  \;\;\infty[, \;\;\;y_{\star} \in I_y = [\frac{1}{\sqrt{2}}, \;\;1 [,
\end{equation}
and are related to one another by
\begin{equation}
y_{\star} = F_{1,\star}(\zeta), \;\;\;\zeta = G_{1,\star}(y_{\star}). \label{yzeta}
\end{equation}
The functions $F_{1,\star}$ and $G_{1,\star}$ are defined by Equations~(\ref{F1aG1a1}) and (\ref{F1aG1a}). When
the wavelength,  $\lambda$, and the period, $T$, are known from observations, or alternatively when the phase velocity, $v_{\rm ph}$, is known, then  there is infinite number of
couples $(\zeta, y_{\star})$  that reproduce the observations. We can let
$\zeta$ vary over the interval $[1, ;\;\;\infty[$ and compute for
each value of $\zeta$ the corresponding value of $y_{\star} = F_{1,\star}(\zeta)$,
or conversely, let $y_{\star}$ vary over the interval $[ 1/\sqrt{2}, \;\; 1 [$ and compute for each value of $y_{\star}$ the corresponding
value of $\zeta = G_{1,\star}(y)$.

\section{Asymptotic analytic expressions for the damping length.}

Let us now turn to the damping length. In order for the kink MHD waves to be damped by resonant absorption additional physics has to be introduced in the equilibrium model. As explained by \citet{terradas2010} the required additional physics is non-uniformity of the local Alfv\'{e}n velocity. For a constant magnetic field this implies a non-uniform density. \citet{terradas2010} derived an asymptotic expression for the damping length.  This asymptotic expression is derived in the approximation that the non-uniform layer is thin.  This is the so-called thin boundary approximation. In what follows we shall refer to it as the TB approximation. The true density discontinuity is replaced by a continuous variation of density in the interval $] R - \frac{l}{2} \;\; R + \frac{l}{2}[$. $l$ is the density inhomogeneity length scale. The use of the TB approximation results in the mathematical simplification that the MHD waves can be solutions for uniform plasmas that are connected over the dissipative layer by jump conditions. In this way we can avoid  solving the non-ideal MHD wave equations. The jump condition for the ideal Alfv\'{e}n singularity was introduced on an intuitive manner by \citet{hollweg1988}  and put on a firm mathematical basis by \citet{sakurai1991}, \citet{goossens1995}, and  \citet{goossensruderman1995} for the driven problem and by \citet{tirry1996} for the eigenvalue problem. Jump conditions were discussed and used in, e.g., \citet{goossens2006,goossens2011} and \citet{goossens2008rev}. In case of  the TB approximation combined with the TT approximation the inclusion of the effect of a non-uniform layer in the dispersion relation is relatively simple. Dispersion relation (Equation~(\ref{TrueDisc1})) is now modified as 
\begin{eqnarray}
\rhoi (\omega^2 - \omega_{\rm A,i}^2) &+& \rhoe (\omega^2 - \omega_{\rm A,e}^2) = \nonumber \\
&&i  \pi \frac{ m/r_{\rm A}}{\rho(r_{\rm A}) \left| \Delta_{\rm A}\right|} \rhoi (\omega^2 - \omega_{\rm A,i}^2) \rhoe (\omega^2 - \omega_{\rm A,e}^2).
\label{DRelationTTTB}
\end{eqnarray}
In Equation~(\ref{DRelationTTTB}) $r_{\rm A}$ denotes the position of the Alfv\'{e}n resonance. In the TB approximation it is natural to adopt $r_{\rm A} = R$ but see the comment by \citet{soler2009}. The use of the jump conditions is not restricted to thin non-uniform layers as can be seen from, e.g., \citet{tirry1996}. The solution of Equation~(\ref{DRelationTTTB}) is now a complex wavenumber $k_z = k_{\rm R} + i k_{\rm I}$. Its imaginary part $k_{\rm I}$ reflects the spatial damping of the wave.  \citet{terradas2010} solved Equation~(\ref{DRelationTTTB}) in the assumption that $k_{\rm I} << k_{\rm R}$, i.e. that damping is weak.  In that case the real part $k_{\rm R}$ is given by Equation~(\ref{TrueDisc2}), i.e., $k_{\rm R} = k_{\star}$. The result of the  analysis  for $k_{\rm I}$ is 
\begin{equation}
\frac{k_{\rm I}}{k_{\star}} = \frac{\pi}{8}\frac{m}{R} \frac{ (\rhoi - \rhoe)^2}{ \rhoi + \rhoe} \frac{1}{ \left| \der \rho / \der r \right|_{r_{\rm A}}}.
 \label{kI1}
\end{equation}
The damping length $L_{\rm D}$ is defined as $L_{\rm D} = 1/k_{\rm I}$.  Equation~(\ref{kI1}) can be rewritten as equation for the damping length over wavelength, namely
\begin{equation}
\frac{L_{\rm D}}{\lambda} =  \frac{4}{\pi^2} \frac{ R}{m} \frac{\rhoi + \rhoe}{ (\rhoi- \rhoe)^2} \left| \frac{\der \rho}{ \der r} \right|_{r_{\rm A}}.
 \label{LD1}
\end{equation}
Note that the effect of the inhomogeneous layer is contained in the value of the spatial derivative of density at the resonant position. Note also  Equation~(\ref{LD1}) is valid for all $m\geq  1$. In what follows we concentrate on kink waves with $m=1$.  It makes sense to adopt
\begin{equation}
 \left| \frac{\der \rho}{ \der r} \right|_{r_{\rm A}} = \alpha \frac{ \rhoi - \rhoe}{l}
\label{DensityDerivative}
\end{equation}
$\alpha$ is a numerical factor that depends on the profile of the variation of density in the inhomogeneous layer. For a sinusoidal density profile $\alpha = \pi /2$, while for a linear profile $\alpha = 1$.  With Equation~(\ref{DensityDerivative}) the expression in Equation~(\ref{LD1}) is now \citep[see also Equation~(13) of][]{terradas2010}
\begin{equation}
\frac{L_{\rm D}}{ \lambda} =  F \frac{ 1}{ l / R} \frac{\rhoi + \rhoe}{ \rhoi- \rhoe}, 
\label{LD2}
\end{equation}
where $F = 4 \alpha /\pi^2$ is a numerical factor. $F$ has $\alpha$ in it. For a sinusoidal profile $F = 2 /\pi$ and for a linear profile  $F= 4/\pi^2$. In what follows we shall adopt the value $F = 2/\pi$ for a sinusoidal profile.  \citet{terradas2010} have gone beyond the TB and the TT approximation in an attempt to assess the accuracy of the analytical expression of Equation~(\ref{LD2}). Their conclusion was that Equation~(\ref{LD2}) is accurate far beyond its interval of applicability.  On another note we would like to point out  that Equation~(\ref{LD2}) is obtained  from an asymptotic analysis of the boundary value problem. This asymptotic analysis adopts an exponentially decaying solution. In principle this exponentially decaying solution is only valid for $z \rightarrow \infty$ in the same way as the exponentially decaying solution in time is strictly only valid for $t \rightarrow \infty$. However, numerical analysis by \citet{pascoe2012} shows that deviations from the exponential damping only occur for relatively small density contrasts and short distances.

At this point it is important to note that in the TT and TB approximations the analysis for temporal damping of standing waves and spatial damping of propagating waves is equivalent. The expressions for $\tau_{\rm D}/ T$ and $L_{\rm D} /\lambda$ are exactly the same. A consequence of this fact is that we can repeat Section~3 of \citet{goossens2008} with the period $T$ and damping time $\tau_D$ replaced with wavelength $\lambda$ and damping length $L_{\rm D}$.  We rewrite Equation~(\ref{LD2})  in terms of $\zeta$, namely
\begin{equation}
\frac{ L_{\rm D}}{\lambda} = \frac{2}{\pi} \frac{\zeta + 1}{\zeta - 1} \frac{1}{l/R}. \label{LD3}
\end{equation}
This is the second key equation of the present investigation. Let us now look at Equation~(\ref{LD3}) from a seismic point of view. If we have observed values  of the wavelength  $\lambda$ and the damping length $L_{\rm D}$ and we convince ourselves that Equation~(\ref{LD3}) is a good first analytical approximation of the damping length   then we can invert Equation~(\ref{LD3}) for either $\zeta$ or $l/R$. Actually we shall do both. Let us first solve Equation~(\ref{LD3}) for $\zeta$ and find
\begin{equation}
\zeta = \frac{\frac{l}{2 R} \frac{\pi L_{\rm D}}{\lambda} + \;1}{\frac{l}{2 R} \frac{\pi L_{\rm D}}{\lambda} - 1}.\label{zeta2}
\end{equation}
The reader must keep in mind that we are looking at this equation from a seismic point
of view. Hence wavelength  and damping length can be considered as known
from observations and it makes sense to denote $\pi L_{\rm D}/ \lambda$ as a
constant $C$. In addition it is convenient to abbreviate $l/2R$ as
$x$. Hence
\begin{equation}
x = \frac{l}{2 R}, \qquad C = \frac{\pi L_{\rm D}}{\lambda}. \label{z}
\end{equation}
Equation~(\ref{zeta2}) can be rewritten as
\begin{equation}
\zeta = \frac{C x +1}{C x -1}.
\label{zeta3}
\end{equation}
Here are several observations to be made. First, since $\zeta >0$ it follows from Equation~(\ref{zeta3}) that for a given ratio $L_{\rm D}/\lambda$ there is a lower  bound for the inhomogeneity length scale, namely
\begin{equation}
x = \frac{l}{2 R} > \frac{\lambda}{\pi L_{\rm D}} = \frac{1}{C} = x_{\rm min}. \label{zmin}
\end{equation}
Second, realize that Equation~(\ref{zeta3}) is a parametric representation of $\zeta$ in terms of $x= l/2R$. In order to make this point very explicit, we introduce the function $G_{2,\star}$ defined as
\begin{equation}
G_{2,\star}\;:\;]\frac{\displaystyle 1}{\displaystyle C}, \;\;\; 1]
\;\rightarrow \mathbb{R},\;\; x \; \Rightarrow \; G_2(x) =
\frac{C x + 1}{C x - 1}. \label{G2}
\end{equation}
It is easy that show that
\begin{equation}
\frac{\der G_{2,\star}}{\der x} = \frac{- 2 C}{(C x - 1)^2} <0.
\end{equation}
Hence $\zeta$ is a decreasing function of $x = l/2R$. It attains
its absolute minimum for $x = l/2R = 1$. This minimal value is
\begin{equation}
\zeta_{\rm min} = G_{2,\star}(1) =  \frac{C +1}{C - 1}.
 \label{zetamin}
\end{equation}
Conversely $\zeta$ attains its maximal value in the limit $x \rightarrow 1/C$, namely 
\begin{equation}
\lim_{z \rightarrow 1/C} G_{2,\star}(x) = \infty.
\end{equation}
With this information on the function $G_{2,\star}$ we can refine its
definition (Equation~(\ref{G2})) as follows
\begin{equation}
G_{2,\star}\;:\;]\frac{1}{C},\; 1] \;\rightarrow\;[\frac{C + 1}{C - 1}, \;\;\infty[,\;\; x \; \Rightarrow \; G_{2,\star}(x) = \frac{C x + 1}{C x - 1}. \label{G2a}
\end{equation}
With the minimal value for $\zeta$ (Equation~(\ref{zetamin})) we can slightly
improve on the upper  bounds for $y_{\star}$ and  for $v_{\rm A,i}$ as 
\begin{equation}
y_{\star} \leq  \left(\frac{C}{C+1} \right)^{1/2},\qquad v_{\rm A,i} \leq \frac{\lambda}{T} \left(\frac{C}{C + 1} \right)^{1/2} = v_{\rm ph}\left(\frac{C}{C + 1} \right)^{1/2}.
\label{vAi2}
\end{equation}

With the help of the information on the bounds for $y_{\star}$ and $\zeta$
we can refine the definitions given in  Equations~(\ref{F1aG1a1}) and (\ref{F1aG1a}) for $F_{1,\star}$ and $G_{1,\star}$,
respectively, to their  final versions as
\begin{eqnarray}
F_{1,\star} \;& :  &\;[\frac{C + 1}{C - 1}, \;\;\infty[ \;\rightarrow ]\frac{1}{\sqrt{2}},\;\; \left(\frac{C}{C +1 } \right)^{1/2} ] ,\nonumber \\
 &&\zeta  \Rightarrow  F_{1,\star}(\zeta) = \frac{1}{\sqrt{2}} \left( \frac{\zeta +1 }{\zeta}\right)^{1/2}, \label{F1G11}\\
G_{1,\star} \;& : &\; ]\frac{1}{\sqrt{2}} , \;\;\;\left(\frac{ C +1}{ C } \right)^{1/2}  \;\; ]\;\;\rightarrow [\frac{C + 1}{C - 1}, \;\;\infty[ ,\nonumber \\
&& y_{\star} \Rightarrow  G_{1,\star}(y_{\star}) = \frac{1}{2 y_{\star}^2 - 1}. \label{F1G1}
\end{eqnarray}

Let us now solve  Equation~(\ref{LD3}) for $x= l/2R$ and find
\begin{equation}
x = \frac{l}{2 R} = \frac{1}{C} \frac{\zeta + 1}{\zeta - 1}. \label{z2}
\end{equation}
Equation~(\ref{z2}) is a parametric representation of $x = l/2R$ in terms
of $\zeta$. As before we make this point explicit by introducing the
function $F_{2,\star}$ defined as
\begin{equation}
F_{2,\star} \;:\;[\frac{C + 1}{C - 1}, \;\;\;
\infty [ \;\rightarrow \mathbb{R},\;\; \zeta \; \Rightarrow \;
F_{2,\star}(\zeta) = \frac{1}{C }\frac{\zeta + 1}{\zeta - 1}. \label{F2}
\end{equation}
Since
\begin{equation}
\frac{\der F_2}{\der \zeta} = \frac{- 2 }{ C  (\zeta - 1)^2} <0
\end{equation}
$F_{2,\star}(\zeta)$ is a decreasing function of $\zeta$. In addition
\begin{equation}
F_{2,\star}(\zeta_{\rm min}) = 1, \qquad \lim_{\zeta \rightarrow \infty} F_{2,\star}(\zeta) = \frac{1}{C}.
\end{equation}
With this information on the function $F_{2,\star}$ we can refine its definition (Equation~(\ref{F2})). Combined with the definition for
the function $G_{2,\star}$ (Equation~(\ref{G2a})) we obtain
\begin{eqnarray}
F_{2,\star} \;& : & \;[\frac{C + 1}{C - 1}, \;\;\; \infty [ \;\rightarrow ]\frac{1}{C}, \;\;\; 1],\;\; \zeta \; \Rightarrow \; F_{2,\star}(\zeta) =  \frac{1}{C} \frac{\displaystyle \zeta + 1}{\displaystyle \zeta - 1}, \nonumber \\ \label{F2G21} \\
G_{2,\star}\;& : &\;]\frac{1}{C}, \;\;\; 1] \;\rightarrow\;[\frac{C + 1}{C - 1}, \;\;\infty[,\;\; x \; \Rightarrow \; G_{2,\star}(x) = \frac{C x + 1}{C x - 1}. \nonumber \\ \label{F2G2}
\end{eqnarray}
$F_{,\star}2$ is the inverse function of $G_{2,\star}$, i.e., $F_{2,\star} = G_{2,\star}^{-1}$ and
conversely $G_{2,\star}$ is the inverse function of $F_{2,\star}$.

\section{Analytical seismology}

\subsection{Summary}

Let us recapitulate the key results of the previous section. The three quantities that we assume to be known from observations are the wavelength, $\lambda$, the damping length, $L_{\rm D}$, and the period, $T$. Alternatively, we can use the phase velocity, $v_{\rm ph}$, instead of the period, $T$, if $v_{\rm ph}$ is known instead of $T$. Analytical theory based on the TT and TB approximations gives us two equations, namely  Equations~(\ref{lambda})  and (\ref{LD3})  that express the wavelength  $\lambda$ and the damping length $L_{\rm D}$ in terms of the density contrast, $\zeta$, the normalized Alfv\'{e}n velocity,  $y_{\star} = v_{\rm A,i} \;T/ \lambda = v_{\rm A,i} / v_{\rm ph}$, and the inhomogeneity length scale normalized to the radius of the tube, $x = l/2R$. These three quantities $\zeta$, $y_{\star}$, and $x$ are the seismic quantities in the sense that they are the quantities that we aim to determine with the use of observed values of the wavelength, $\lambda$, the damping length, $L_{\rm D}$, and either the period, $T$, or the phase velocity, $v_{\rm ph}$. Since we have only two equations that relate the three unknown quantities to the three observed quantities there are an infinite number of solutions. The seismic variables are constrained to the following intervals
\begin{eqnarray}
\zeta &\; \in \; & I_{\zeta} = \left[\frac{C + 1}{C - 1}, \;\;\infty\right[  \label{yzetaz10}\\
y_{\star} & \; \in \; & I_{y_{\star}} = \left]\frac{1}{\sqrt{2}}, \, \, \left(\frac{C}{C +1 } \right)^{1/2}\right]  \\
x & \;\in \; & I_x = \left[\frac{1}{C}, \;\;1\right], \label{yzetaz1}
\end{eqnarray} 
and are related to one another by
\begin{eqnarray}
y_{\star} &\; =\; & F_{1,\star}(\zeta), \;\;\; \zeta \; =\;  G_{1,\star}(y_\star), \label{yzetaz21}\\
x & \; =\; & F_{2,\star}(\zeta),\;\;\; \zeta  \; =\; G_{2,\star}(x) \label{yzetaz22}
\end{eqnarray}
The functions $F_{1,\star}$, $G_{1,\star}$, $F_{2,\star}$, and $G_{2,\star} $ are defined by Equations~(\ref{F1G11}), (\ref{F1G1}), (\ref{F2G21}) and
(\ref{F2G2}), respectively.

In Equations~(\ref{yzetaz21}) and (\ref{yzetaz22}) only two equations are independent since $G_{1,\star}$ is the inverse function of $F_{1,\star}$ and $G_{2,\star}$ is the inverse function of $F_{2,\star}$. Equations~(\ref{yzetaz21}) and (\ref{yzetaz22}) give us the infinitely many solutions of the seismic inversion in parametric form. Each of the three unknowns can be used as parameter and the two remaining unknowns can be expressed in terms of that parameter. For example choose $\zeta$ as parameter. Let $\zeta$ take on all values in $I_{\zeta}$ and compute the corresponding values of $y_{\star}$ and $x$ by the use of $y_\star = F_{1,\star}(\zeta)$ and $z = F_{2,\star} (\zeta)$. Or choose $y_{\star}$ as parameter. Let $y_{\star} $ take on all values in $I_{y_\star}$ and then compute the corresponding values of $\zeta$ and $x$ by the use of $\zeta = G_{1,\star}(y_\star)$ and $z = F_{2,\star}(G_{1,\star}(y_\star))$.  Finally, use $x$ as parameter to define the solutions of the inversion problem. Let $x$ take on all values in $I_x$ and then compute the corresponding values of $y_{\star}$ and $\zeta$ by the use of $\zeta = G_{2,\star}(x)$ and $y_{\star} = F_{1,\star}(G_{2,\star}(x))$.

Note that instead of the seismic variables $y_\star$, $x$, and $\zeta$, the inversion can be performed for the variables $v_{\rm A,i}$, $l/R$, and $\zeta$, which have a more obvious physical meaning.  The relation between both sets of seismic variables is
\begin{equation}
 v_{\rm A,i} = y_\star \frac{\lambda}{T} = y_\star v_{\rm ph}, \qquad \frac{l}{R} = 2 x, \qquad \zeta = \zeta.
\end{equation}
 These alternative seismic variables are constrained to the intervals
\begin{eqnarray}
v_{\rm A,i} & \; \in \; & I_{v_{\rm A,i}} = \left]\frac{v_{\rm ph}}{\sqrt{2}}, \,\, v_{\rm ph}\left( \frac{C}{C + 1} \right)^{1/2}\right], \label{intva} \\
\frac{l}{R} & \;\in \; & I_{l/R} = \left[\frac{2}{C}, \;\;2\right], \\
\zeta &\; \in \; & I_{\zeta} = \left[\frac{C + 1}{C - 1}, \;\;\infty\right[. \label{intzeta}
\end{eqnarray} 

\subsection{Warning}

The present inversion scheme for propagating MHD waves, as its twin version for standing MHD waves, is simple to use. However, caution is required. First of all the analytic expression used in the inversion scheme for the damping length or conversely for the damping time states that the damping length or damping time are inversely proportional to $l/R$. This relation is definitely very accurate when the non-uniform layers are sufficiently thin. When these analytical expression are applied to fully non-uniform wave guides then the prediction would be that the waves undergo extremely fast damping. This result might be erroneous. Equation~(\ref{LD3}) is derived for a thin non-uniform transitional layer and the kink MHD wave is essentially a surface Alfv\'{e}n wave. When the non-uniform layers become very thick the MHD waves are no longer surface waves \citep[see][]{arregui2006,vandoorsselaere2007}. Their damping deviates from that for surface Alfv\'{e}n waves and can be quite different from that predicted by  Equation~(\ref{LD3}). So it is wise to stay away from high values of $l/R$. 

Secondly we  neglected non-uniformity along the wave guide. For standing MHD waves longitudinal non-uniformity in both density \citep[see, e.g.,][]{andries2005,arregui2005} and magnetic field \citep[see, e.g.,][]{verth2008,ruderman2008} affects the periods and the ratios of periods, but the damping by resonant absorption is not affected by longitudinal stratification \citep[see, e.g.,][]{andries2005,arregui2005,dymova2006}. Also longitudinal stratification does not cause additional damping or amplification. As the periods are concerned a longitudinally  averaged density can be defined that produces the same period  in a loop that is homogeneous in the longitudinal direction. For propagating waves the story is different. Radial stratification causes resonant absorption and damping but longitudinal stratification has an effect on the amplitude of the wave as shown by \citet{soler2011c}.  Longitudinal stratification causes the amplitude of the wave to increase with height and might partially or fully hide the damping due to resonant absorption. If the effect of longitudinal stratification is not removed from the observations, then Equation~(\ref{LD3}) uses and underestimate of the damping length due to resonant absorption and this will affect the inversion result. 

Finally, we recall that the present inversion scheme is based on linear theory. \citet{ruderman2010} showed that nonlinearity can strongly increase the efficiency of damping due to resonant absorption. Although linear theory is accurate enough to describe the small-amplitude waves observed in the solar corona  \citep[e.g.,][]{tomczyk2007}, the influence of nonlinearity may be important for kink waves propagating in other structures in the solar atmosphere. The conditions for which nonlinear effects become important are discussed in \citet{ruderman2010}.

\section{Example }

As an illustrative example of the technique described in the previous Sections we re-analyze the CoMP observations of running coronal waves \citep{tomczyk2007,tomczyk2009}. \citet{verth2010} showed that CoMP observations are  consistent with an interpretation based on resonantly damped propagating kink waves. Here we perform the full seismological inversion using the CoMP data.

We use $v_{\rm A,i}$, $l/R$, and $\zeta$ as our seismic variables. To perform the analytic inversion,  we need observational values of the wavelength, $\lambda$, the damping length, $L_{\rm D}$, and the period, $T$. Alternatively, in the inversion of $v_{\rm A,i}$ we can use the phase velocity, $v_{\rm ph}$, instead of the period, $T$, because both quantities are related by the wavelength (Equation~(\ref{eq:periodvph})). Since the observational value of $v_{\rm ph} = 0.6$~Mm~s$^{-1}$ is provided by \citet{tomczyk2009}, we use $v_{\rm ph}$ in the inversion scheme.

Next we compute the parameter $C = \pi L_{\rm D} / \lambda$ using observed quantities. In principle  we need observational values of both the wavelength, $\lambda$, and the damping length, $L_{\rm D}$. However, we note that what we actually need is an observational estimation of the ratio $L_{\rm D} / \lambda$.  The theoretical expression of $L_{\rm D} / \lambda$ is given in Equation~(\ref{LD3}). Both $\lambda$ and  $L_{\rm D}$ depend on the wave period, $T$, but their ratio is independent of $T$.  \citet{verth2010} took advantage of this result and expressed the ratio $L_{\rm D} / \lambda$ in terms of the constant parameter $\xi_{\rm E}$ as
\begin{equation}
 \frac{L_{\rm D}}{\lambda} = \xi_{\rm E},
\end{equation}
that allows us to rewrite $C$ as
\begin{equation}
 C = \frac{\pi L_{\rm D}}{\lambda} =  \pi \xi_{\rm E}.
\end{equation}
The parameter $\xi_{\rm E}$ contains all the information about the properties of the waveguide. After performing a frequency fit to the CoMP data \citet{verth2010} found that the best estimate is $\xi_{\rm E} = 2.69$, which corresponds to $C = 8.45$. Note that the fit of \citet{verth2010} was made for the wave path represented with a dashed line in Figure~1 of \citet{tomczyk2009}.  The spatial resolution of CoMP observations was not enough  to  isolate individual coronal loops, i.e., indivitual waveguides, and the signal was spatially averaged inside the dotted region enclosing the dashed line \citep[see details in][]{verth2010}. Since the power was averaged inside the dotted region \citet{verth2010} obtained an averaged value of $\xi_{\rm E}$. For this reasion, the seismic variables inverted using this averaged value of $\xi_{\rm E}$ have to be interpreted as averaged values as well.  The estimated $\xi_{\rm E}$ using the frequency fit in \citet{verth2010} is consistent with previous estimations using TRACE observations, which points out the validity of the analysis. We stress again the ratio $L_{\rm D}/\lambda$ is enough for the inversion. However if  information of both  $\lambda$ and $L_{\rm D}$ is available in the observations, it can be used to directly compute $C$.

Now we use Equations~(\ref{intva})--(\ref{intzeta}) we compute the intervals of the seismic variables $v_{\rm A,i}$, $l/R$, and $\zeta$, namely
\begin{eqnarray}
 v_{\rm A,i} &\; \in \; &  ]424, \, \,  567] \,\,\textrm{km s}^{-1}, \\
\frac{l}{R} &\; \in \; &  [0.24, \, \, 2], \\
\zeta  &\; \in \; & [1.27, \, \, \infty [.
\end{eqnarray}
The variable that can be constrained in the narrower range is $v_{\rm A,i}$, whereas $\zeta$ remains in practice unconstrained. The lower value of $l/R$ is imposed by the observations while the upper value is imposed by the model.  

\begin{figure}
	\centerline{\includegraphics[width=.9\columnwidth]{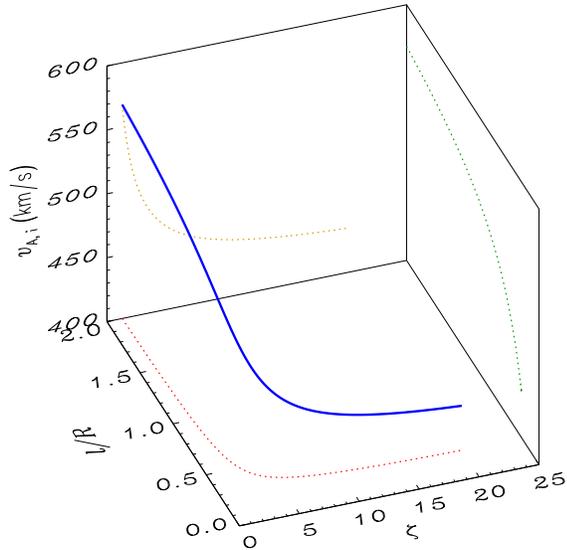}}
	\caption{Result of the analytic inversion scheme applied to CoMP observations of propagating coronal waves. The solid line is the 1D solution in the 3D space of parameters $v_{\rm A,i}$--$l/R$--$\zeta$. The dotted lines are projections of the full solution to the various planes.}
	\label{fig:seismology}
\end{figure}

Now we perform the full inversion and use $\zeta$ as a free variable. We compute the corresponding values of $v_{\rm A,i}$ and $l/R$ that are compatible with the observations. To do so we use the relations given in Equations~(\ref{yzetaz21}) and (\ref{yzetaz22}), namely
\begin{eqnarray}
 v_{\rm A,i} &=& v_{\rm ph}F_{1,\star}(\zeta) =  \frac{v_{\rm ph}}{\sqrt{2}}\left( \frac{\zeta + 1}{\zeta} \right)^{1/2}, \\
\frac{l}{R} &=& 2 F_{2,\star}(\zeta) = \frac{2}{C} \frac{\zeta+1}{\zeta-1}.
\end{eqnarray}
The corresponding solutions form a 1D curve in the 3D space of parameters $v_{\rm A,i}$--$l/R$--$\zeta$ (Figure~\ref{fig:seismology}). Any point on this curve is equally compatible with the observations.

\section{Conclusion} 

In this paper we have presented an analytical seismological inversion scheme for propagating MHD waves in the solar atmosphere. This scheme uses the observational information on wavelengths  and  damping lengths in a consistent manner  and  is based on  approximate asymptotic expressions for the theoretical values of both quantities. The scheme also needs observational values of the wave periods or, alternatively, of the phase velocities. The seismological inversion scheme for propagating waves shown here is the counterpart to that developed by \citet{goossens2008} for standing waves.

Using the Alfv\'en velocity, the inhomogeneity length scale, and the density contrast as seismic variables, we have shown that they can be constrained in intervals which depend on the observed values of wavelength, damping length, and phase velocity. The seismic variable that can be constrained the most is the Alfv\'en velocity, while the density contrast is the less constrained variable. The whole collection of values of the seismic variables that are compatible with the observations describe a 1D curve in the 3D space of variables. A priori, with no additional information on the uncertainties of the observed quantities, any point on this curve can equally explain the observations. As described in  \citet{arregui2011}  in the case of the inversion scheme for standing waves \citep{goossens2008}, additional information can be consistently implemented in the Bayesian framework, which results in the full determination of the three unknowns, with correctly propagated uncertainties. As for standing waves, the use of Bayesian analysis with the present scheme for propagating waves is possible.

The inversion scheme presented here is based on a simple model for the magnetic wave guide. It is a challenge for future works to incorporate more realistic ingredients to the model in order to determine their impact on the seismological inversion. In particular, the effects of flow \citep[see, e.g.,][]{terradasflow2010,soler2011b}, longitudinal stratification \citep[see, e.g.,][]{soler2011c}, and nonlinearity \citep[see, e.g.,][]{ruderman2010} are worth being explored in forthcoming works.

\acknowledgements{This research was begun when M.G. was a visitor of the Solar Physics
Group at the UIB.  It is pleasure for M.G. to acknowledge the warm
hospitality of the Solar Physics Group at the UIB. M.G. acknowledges support from KU Leuven via GOA/2009-009. R.S. acknowledges support from a Marie Curie Intra-European Fellowship within the European Commission 7th Framework Program  (PIEF-GA-2010-274716).  I.A. acknowledges support by a Ram\'on y Cajal Fellowship by the Spanish Ministry of Economy and Competitiveness (MINECO). J.T. acknowledges support by a Ram\'on y Cajal Fellowship by the MICINN.  R.S. and J.T. acknowledge support from CAIB through the `grups competitius' scheme and FEDER Funds. All the authors acknowledge the support from the Spanish MICINN/MINECO and FEDER funds through project AYA2011-22846.}

\bibliographystyle{apj} 
\bibliography{refs}

\end{document}